\documentclass[aps,prl,twocolumn,showpacs,superscriptaddress]{revtex4-1}
\usepackage{graphicx}
\usepackage{amsmath}
\usepackage{amssymb}
\usepackage{colordvi}
\usepackage{mathrsfs}
\usepackage{bm}
\usepackage{verbatim}
\usepackage{dcolumn}
\usepackage{bm}
\usepackage{epsfig}
\usepackage{subfigure}

\begin{document}
\title{Nonmagnetic-Doping Induced Quantum Anomalous Hall Effect in Topological Insulators}
\author{Shifei Qi}
\affiliation{ICQD, Hefei National Laboratory for Physical Sciences at Microscale, CAS Key Laboratory of Strongly-Coupled Quantum Matter Physics, and Department of Physics, University of Science and Technology of China, Hefei, Anhui 230026, China.}
\affiliation{Department of Physics, Hebei Normal University, Shijiazhuang, Hebei 050024, China}
\author{Ruiling Gao}
\affiliation{Institute of Materials Science, Shanxi Normal University, Linfen, Shanxi 041004, China}
\affiliation{Department of Physics, Hebei Normal University, Shijiazhuang, Hebei 050024, China}
\author{Maozhi Chang}
\affiliation{Institute of Materials Science, Shanxi Normal University, Linfen, Shanxi 041004, China}
\affiliation{Department of Physics, Hebei Normal University, Shijiazhuang, Hebei 050024, China}
\author{Tao Hou}
\affiliation{ICQD, Hefei National Laboratory for Physical Sciences at Microscale, CAS Key Laboratory of Strongly-Coupled Quantum Matter Physics, and Department of Physics, University of Science and Technology of China, Hefei, Anhui 230026, China.}
\author{Yulei Han}
\affiliation{ICQD, Hefei National Laboratory for Physical Sciences at Microscale, CAS Key Laboratory of Strongly-Coupled Quantum Matter Physics, and Department of Physics, University of Science and Technology of China, Hefei, Anhui 230026, China.}
\author{Zhenhua Qiao}
\email[Correspondence author:~~]{qiao@ustc.edu.cn}
\affiliation{ICQD, Hefei National Laboratory for Physical Sciences at Microscale, CAS Key Laboratory of Strongly-Coupled Quantum Matter Physics, and Department of Physics, University of Science and Technology of China, Hefei, Anhui 230026, China.}

\date{\today{}}

\begin{abstract}
  Quantum anomalous Hall effect (QAHE) has been experimentally observed in magnetically doped topological insulators. However, ultra-low temperature (usually below 300 mK), which is mainly attributed to inhomogeneous magnetic doping, becomes a daunting challenge for potential applications. Here, a \textit{nonmagnetic}-doping strategy is proposed to produce ferromagnetism and realize QAHE in topological insulators. We numerically demonstrated that magnetic moments can be induced by nitrogen or carbon substitution in Bi$_2$Se$_3$, Bi$_2$Te$_3$, and Sb$_2$Te$_3$, but only nitrogen-doped Sb$_2$Te$_3$ exhibits long-range ferromagnetism and preserve large bulk band gap. We further show that its corresponding thin-film can harbor QAHE at temperatures of 17-29 Kelvin, which is two orders of magnitude higher than the typical temperatures in similar systems. Our proposed \textit{nonmagnetic} doping scheme may shed new light in experimental realization of high-temperature QAHE in topological insulators.
\end{abstract}

\maketitle
Quantum anomalous Hall effect (QAHE)~\cite{Haldane} is the counterpart of quantum Hall effect in the absence of Landau levels from strong magnetic field. Due to the topological protection of spatial separation, the chiral edge modes of QHE/QAHE is believed to have immense application potential in future dissipationless quantum devices~\cite{he2018topological}. Since its initial predication in a honeycomb-lattice model~\cite{Haldane}, numerous effort has been made to exploring new platforms for realizing QAHE in related material systems~\cite{proposal1,proposal2,proposal3,proposal4,proposal5,proposal51,proposal52,proposal6,proposal7,proposal8,proposal9,proposal10,proposal11,proposal12}. Because of the virtue of inherent strong spin-orbit coupling in $\mathbb{Z}_2$ topological insulators (TIs)~\cite{TopologicalInsulator1,TopologicalInsulator2}, it has narrowed the search for suitable materials possessing intrinsic magnetism. One natural approach to induce magnetism in TIs is to dope magnetic elements~\cite{DMS,proposal4,TI-magnetism1,TI-magnetism2,TI-magnetism3}. This has indeed resulted in the first experimental realization of QAHE in Cr/V-doped (Bi,Sb)$_2$Te$_3$ thin films~\cite{ChangCuiZu,ExperimentalQAHE1,ExperimentalQAHE2,2015NatMat}. So far, all the experimentally observed QAHEs in doped TIs were achieved at extremely low temperatures, typically lower than 300 mK, making drastically increase the observation temperature a daunting challenge for potential applications.

In magnetically-doped (Bi,Sb)$_2$Te$_3$ thin films, inhomogeneous ferromagnetism is regarded as one of the main factors that leads to the unexpected low QAHE observation temperature~\cite{PNAS}. Another possible factor is the appearance of the dissipative conduction channels due to metallization of the magnetically doped TI's bulk-like region~\cite{APL}. Empirically, codoping is an effective way to improve ferromagnetic order, as confirmed in dilute magnetic semiconductors (DMSs)~\cite{DMS}. Our theoretical study has approved that the observed temperature of the QAHE can be largely improved by codoping Sb$_2$Te$_3$ with $p$-type vanadium and $n$-type iodine dopants~\cite{Qi2016}. Indeed, an increased QAHE temperature was observed in Cr and V-codoped (Bi,Sb)$_2$Te$_3$ thin films~\cite{CodopingEXP}. At the optimal Cr/V ratio, full quantization was achieved at 300 mK, an order of magnitude higher than that with single dopants. The obtained Hall hysteresis loop is more square-like, suggesting a reduced magnetic inhomogeneity in Cr- and V-codoped (Bi,Sb)$_2$Te$_3$ thin films~\cite{CodopingEXP}. In the study of DMSs, it is known that homogeneous DMS by magnetic doping is unrealistic~\cite{DMS}. By carefully tuning crystal growth conditions or utilizing codoping method, magnetic inhomogeneity can only be suppressed more or less. This is because strong attractive interaction from doped magnetic impurities are difficult to suppress their aggregation in DMSs~\cite{DMS2}.

Alternatively, many materials by nonmagnetic dopants have exhibited robust high-temperature ferromagnetism. For example, carbon(C)-doped ZnO films deposited by pulsed-laser deposition showed ferromagnetism with Curie temperature higher than 400 K~\cite{CZnO,CZnO2}; Above room-temperature ferromagnetism was also observed in C-doped hexagonal boron nitride~\cite{CBN}; Doping nonmagnetic fluorine in two-dimensional antimonene also realizes robust above-room-temperature ferromagnetism~\cite{Antimonene}. More importantly, no strong attractive interactions exists between nonmagnetic dopants, naturally avoiding the formation of inhomogeneous ferromagnetism and dissipative conduction channels. Therefore, it is of great interest to explore the possibility of realizing high-temperature QAHE in \textit{nonmagnetic}-doped TIs.

In this Letter, we showed that high-temperature QAHE can be realized in nonmagnetically-doped TIs, e.g., by doping nonmagnetic C or N atoms in
Sb$_2$Te$_3$ TI thin films. We took three representative TIs (i.e., Bi$_2$Se$_3$, Bi$_2$Te$_3$, and Sb$_2$Te$_3$) as examples, and found that their magnetic moments can be induced via C or N substitution. For all doping concentrations considered in our study, only N-doped Sb$_2$Te$_3$ can give rise to a long-range ferromagnetic order. Moreover, an estimated Curie temperature of 17-29 Kelvin can be reached at the concentration of 11-22\% N-doped Sb$_2$Te$_3$. Aside from the bulk properties, the band gaps of the corresponding Sb$_2$Te$_3$ thin films can also be as large as 17 meV, much larger than what is needed to sustain ferromagnetic order at the high Curie temperature. Explicit Berry curvature calculation further confirms that the N-doped Sb$_2$Te$_3$ thin film can harbor QAHE. Our proposed nonmagnetic doping approach thus offers a versatile route to realize QAHE in TIs and related materials.

\begin{figure}
  \includegraphics[width=8.5cm,angle=0]{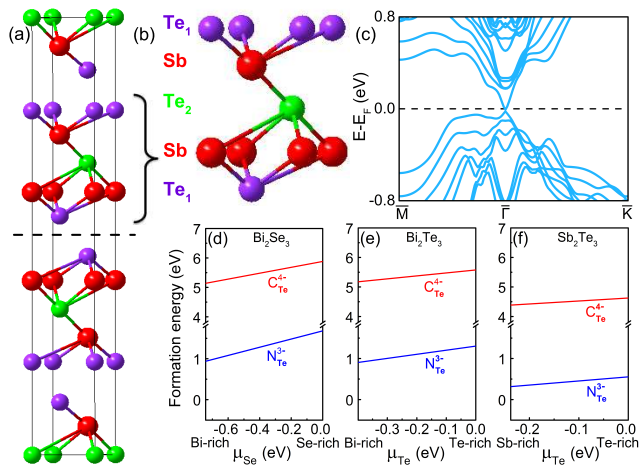}
  \caption{(a) Structure illustration of bulk Sb$_2$Te$_3$. (b) $1\times 1 \times 1$ unitcell of Sb$_2$Te$_3$ with only one of three QLs shown for clarity. (c) Band structure of Sb$_2$Te$_3$ with five QLs. Formation energies of the most stable configurations of single C- or N-doped Bi$_2$Se$_3$ (d), Bi$_2$Te$_3$ (e), and Sb$_2$Te$_3$ (f) as a function of the host element chemical potentials, respectively.}
  \label{Fig1}
\end{figure}

Our first-principles calculations were performed using the projected augmented-wave method~\cite{PAW} as implemented in the Vienna Ab-initio Simulation Package (VASP)~\cite{VASP1}. The generalized gradient approximation (GGA) of Perdew-Burke-Ernzerhof (PBE) type was used to treat the exchange-correlation interaction~\cite{GGA}. $1\times 1\times1$ Bi$_2$Se$_3$, Bi$_2$Te$_3$, and Sb$_2$Te$_3$ cells were chosen in this study. For thin film calculations, the film thickness was chosen to be five quintuple layers (QLs). A vacuum buffer space of 30~\AA  was used to prevent coupling between adjacent slabs. The kinetic energy cutoff was set to 400 eV. During structural relaxation, all atoms were allowed to relax until the Hellmann-Feynman force on each atom is less than 0.01eV/\AA. The Brillouin-zone integration was carried out by using $15 \times 15 \times2$ and $9 \times 9 \times 1$ Monkhorst-Pack grids for bulk and thin film systems, respectively. Unless mentioned otherwise, spin-orbit coupling was considered in all calculations. The Curie temperature $T_{\rm C}$ was estimated within the mean-field approximation $k_{\rm B} T_{\rm C} = \frac{2}{3}xJ$~\cite{CurieTemperature}, where $k_{\rm B}$ is the Boltzmann constant, $x$ is the doping concentration, and $J$ is the exchange parameter obtained from the total energy difference between ferromagnetic and antiferromagnetic configurations at different doping concentrations.

Let us first discuss the possibility of doping N and C atoms in Bi$_2$Se$_3$, Bi$_2$Te$_3$, and Sb$_2$Te$_3$ TIs. As a concrete example, the bulk structure of Sb$_2$Te$_3$ is shown in Fig.~\ref{Fig1}(a). Each QL [see Fig.~\ref{Fig1}(b)] consists of two equivalent Te atoms (Te$_1$), two equivalent Sb atoms, and a third Te atom (Te$_2$). Atoms within one QL are coupled by a strong chemical bond, whereas weak van der Waals forces hold QLs together. To this end, we place the impurities at Se or Te site in $2\times2\times1$ bulk Bi$_2$Se$_3$, Bi$_2$Te$_3$, and Sb$_2$Te$_3$ TIs, and compare their formation energies $\Delta E_{\rm F}$, which can be evaluated by
\begin{eqnarray}\label{eq:erel}
\Delta E_{\rm F}=E_{\rm tot}^{\rm D}-E_{\rm tot}-\sum n_{i} \mu_{i},
\end{eqnarray}
where $E_{\rm tot}^{\rm D}$ is the total energy of the supercell including one nonmagnetic dopant, $E_{\rm tot}$ is the total energy of the supercell, $\mu_{i}$ is the chemical potential for the species $i$ (host atoms or dopants) and $n_{i}$ is the corresponding number that has been added to or removed from the supercell. We chose rhombohedral Bi, rhombohedral Sb, hexagonal Se, hexagonal Te, nonmagnetic body centered-cubic C, antiferromagnetic hexagonal N as reference to evaluate the chemical potentials of the elements. The computational details were adopted from a previous study of magnetically doped TIs~\cite{WGZhu}.

\begin{table}
  \caption{Total magnetic moments (M$_{\rm tot}$) and magnetic moments contributed by C ($M_ {\rm C}$) or N ($M_{\rm N}$) atom in C- or N-doped bulk Bi$_2$Se$_3$, Bi$_2$Te$_3$ , and Sb$_2$Te$_3$. X represents Se or Te atom. The unit of magnetic moment is $\mu_{B}$.}
  \begin{ruledtabular}
  \begin{tabular}{ccccc}
  &\multicolumn{2}{c}{C-doped TIs}&\multicolumn{2}{c}{N-doped TIs}\\ \cline{2-5}
  TIs&X$_{1}$&X$_{2}$&X$_{1}$&X$_{2}$\\ \cline{2-5}
  &$M_{\rm C}$ $M_{\rm tot}$& $M_{\rm C}$ $M_{\rm tot}$& $M_{\rm N}$ $M_{\rm tot}$& $M_{\rm N}$ $M_{\rm tot}$\\ \hline
  Bi$_2$Se$_3$&$0.19$  $0.38$ &$0.00$  $0.00$&$0.62$  $1.00$ &$0.30$  $0.97$ \\
  Bi$_2$Te$_3$&$0.24$  $0.56$ &$0.38$  $1.64$&$0.56$  $1.00$ &$0.18$  $0.69$\\
  Sb$_2$Te$_3$&$0.06$  $0.13$ &$0.00$  $0.00$&$0.00$  $0.00$ &$0.14$  $0.58$\\
  \end{tabular}
  \end{ruledtabular}
  \label{table-1}
\end{table}

As displayed in Figs.~\ref{Fig1}(d)-\ref{Fig1}(f), the formation energies indicate that Se or Te substitutional sites are preferred by N atoms, because N possesses less formation energies (about 1.0 eV) in the whole range of the accessible host element chemical potentials. Particularly for Sb$_2$Te$_3$, the formation energy of N substitution is only in the range of 0.25-0.50 eV. It seems that the estimated formation energy of C substituted Se or Te is much larger than that of N substituted Se or Te, which is about 5.0 eV. While it is noteworthy that C-doped ZnO has already been realized in experiments even if the estimated formation energy of C substituted O in ZnO is about 5.3 eV~\cite{CZnO}. Therefore, one can reasonably conclude that N or C substituted Bi$_2$Se$_3$, Bi$_2$Te$_3$, and Sb$_2$Te$_3$ TIs can be experimentally produced.

\begin{figure}
  \includegraphics[width=8.5cm,angle=0]{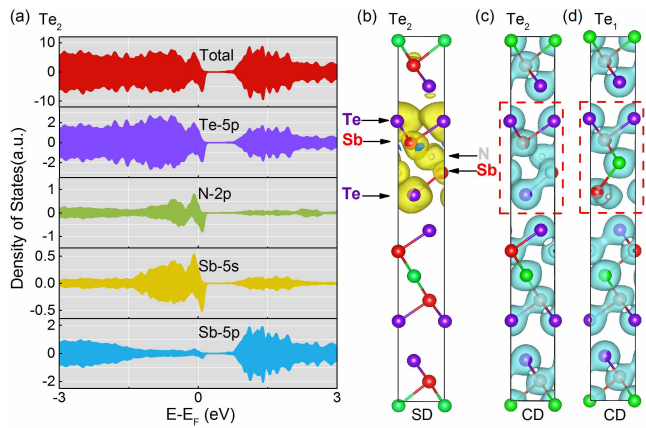}
  \caption{(a) Total and projected density of states for Sb 5p, Te 5p, and N 2p orbitals in N-doped Sb$_2$Te$_3$. (b) Spin density distribution in N-doped Sb$_2$Te$_3$ at Te$_2$ site. Yellow and blue isosurfaces correspond to the majority and minority spin density, respectively. (c-d) Charge density distribution in N-doped Sb$_2$Te$_3$ at Te$_2$ (c) and Te$_1$ (d) sites, respectively.}
  \label{Fig2}
\end{figure}

Next, we focused on whether such nonmagnetic dopants in TIs can induce magnetic moments, and further establish long-range ferromagnetism, as that confirmed in C-doped ZnO~\cite{CZnO,CZnO2}. TABLE~\ref{table-1} summarizes the magnetic moments of C- and N-substituted Bi$_2$Se$_3$, Bi$_2$Te$_3$, and Sb$_2$Te$_3$ TIs. For the C-doped systems, our results show that the magnetic moments can always arise when C substitutes at Te$_1$ or Se$_1$ site, with the corresponding total moments being respectively 0.38, 0.56, and 0.10 $\mu_{B}$ for three different systems; While when C substitutes Te$_2$ site, only the Bi$_2$Te$_3$ system gives a non-vanishing magnetic moment of 1.64 $\mu_{B}$. Compared with C-doping, magnetic moments can always be induced by N doping except the case of N-substitution at Te$_1$ site of Sb$_2$Te$_3$.

The resulted magnetic moments in N- and C-doped TIs can be understood from the calculated local density of states (LDOS), see the typical example of N-doped Sb$_2$Te$_3$ as displayed in Fig.~\ref{Fig2}(a). Strong coupling between 2$p$-orbital of N and Te 5$p$, Sb 5$s$/5$p$ orbitals results in above orbitals near the Fermi level to split along spin directions. The spin-up bands are nearly fully occupied while the spin-down bands are partially filled, leading to a magnetic moment of 0.58 $\mu_{B}$ per unit cell. The spin density distribution in Fig.~\ref{Fig2}(b) further reveals that the ferromagnetism arises from a synergistic effect of the N-dopant and its first-, second-nearest neighboring Sb and Te atoms in one QL. The neighboring Sb and Te atoms are ferromagnetically coupled to the N atom. No $d$ state is observed near the Fermi level, confirming the pure sp-electron ferromagnetism. In addition, little amount of spin polarization along the other direction is observed in one QL. Thus, one can say that the whole QL can be spin-polarized (long-range ferromagnetism) by substituting only one Te atom by one N atom in Sb$_2$Te$_3$. In fact, the magnetic ground state for N-doped Sb$_2$Te$_3$ have been calculated with the inclusion of spin-orbit coupling, and the results indicate that out-of-plane spin polarization is less preferred, which is about 0.5 meV (per unit-cell with one N atom) lower than that of the horizontal axis.

One can also find that no local magnetic moment is induced for N-substitution at the Te$_1$ site of Sb$_2$Te$_3$. As shown in Fig.~\ref{Fig2}(d), it can be attributed to the strong bonding interaction (the bond length is 1.997~\AA) existing between substituted N atom and the nearest Sb atom, which decreases the interaction of the Sb atom and its surrounding Te (Te$_2$ site) atoms in the same QL. While, in the case of N-substitution at Te$_2$ site, the bonding interaction (the bond length is 2.098~\AA) between substituted N atom and its nearest Sb atom is weaker than that in the case of substituting at Te$_1$ site. From Fig.~\ref{Fig2}(c), one can see that this weaker bonding interaction does not decrease the interaction between the substituted N atom and its surrounding Sb atoms, which is in agreement with above observation in Fig.~\ref{Fig2}(a) that ferromagnetism in the case of N-substitution at Te$_2$ arises from a synergistic effect of the N dopant and its first-, second-nearest neighboring Sb and Te atoms within one QL.

\begin{figure}
  \includegraphics[width=8.5cm,angle=0]{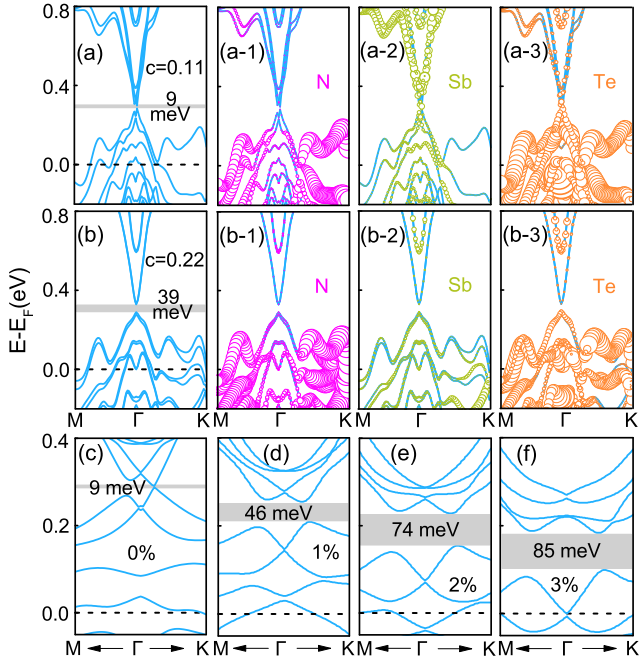}
  \caption{Band structures of N-doped Sb$_2$Te$_3$. (a) At 11\% N-doping concentration, the bulk gap is about 9 meV; (a1)-(a3) Different characters of the bands shown in panel (a), obtained by projecting the Kohn-Sham states onto the local orbitals of a single atom for each element. (b) At 22\% N-doping concentration, the bulk gap enlarges to 39 meV; (b1)-(b3) Different characters of the bands shown in panel (b). (c)Zooming-in of N-doped Sb$_2$Te$_3$ band structure around the band gap. (d)-(f) Band structures of N-doped Sb$_2$Te$_3$ under different in-plane tensile strains. Dashed line denotes the Fermi level.} \label{Fig3}
\end{figure}

Aside from the ferromagnetic ordering of N- and C-doped TIs, two other prerequisites ~\cite{he2018topological} for the observation of QAHE include: (1) The bulk gap exists in N- and C-doped TIs; (2) The surface bandgap can be opened due to the ferromagnetism. To this point, let us continue to investigate the variance of band structures due to the C- and N-doping. First, we focused on the bulk band structures of N- or C-doped Bi$_2$Se$_3$, Bi$_2$Te$_3$, and Sb$_2$Te$_3$ TIs. Figure~\ref{Fig3}(a) displays the bulk band structure of 11\% N-doped Sb$_2$Te$_3$ at Te$_2$ site, opening a small bulk gap about 9 meV. In fact, the bulk gap of N-doped Sb$_2$Te$_3$ can be dramatically increased to 39 meV [See Fig.~\ref{Fig3}(b)] when N-doping concentration increases from 11\% to 22\%. In 11\% N-doped Sb$_2$Te$_3$, further analysis of the characters of different elements [see Figs.~\ref{Fig3}(a-1)-\ref{Fig3}(a-3)] reveals that the valence band (conduction band) close to the band gap is dominated by Te (Sb) atoms, while N contributes mainly to the lower part of the valence bands. These results directly reflect the strong bonding interaction between N and surrounding Sb atoms due to the larger electronegativity of N. If doping concentration increases to 22\%, the contribution from Sb (Te) atoms to conduction (valence band) band close to the gap obviously decreases [see Figs.~\ref{Fig3}(b-1)-\ref{Fig3}(b-3)]. This is because the more N dopants substitute Te atoms, the stronger bonding interaction between N and surrounding Sb atoms arises, and the resulting large energy difference between bonding and anti-bonding orbitals helps enlarge the bulk gap. In addition, our results presented in Figs.~\ref{Fig3}(c)-\ref{Fig3}(f) confirm that appropriate in-plane tensile strains can not only greatly enlarge the bulk gap from 9 mev to 46 (1\% tensile strain), 74 meV (2\% tensile strain), and 85 meV (3\% tensile strain), but also shift the Fermi level closer to the bulk band gap. The strain effect on bulk band gap and Fermi level will be beneficial for realization of high-temperature QAHE in N-doped Sb$_2$Te$_3$. Calculated results for other doping configurations can be found in Supplemental Materials. Unfortunately, no global bulk band gap appears in these systems. Thus, our results prove that only N-doped Sb$_2$Te$_3$ is an ideal candidate for high-temperature QAHE and its corresponding bulk gap can be modulated via doping concentration or applying in-plane strain.

So far, the above results strongly suggest that N-doped Sb$_2$Te$_3$ is a highly potential candidate for exploring high-temperature QAHE. In principle, one still needs to see whether its corresponding thin-film system can also open sizable band gap or not. By analyzing the band structures, we showed that these thin-film systems can indeed support topological states (i.e., QAHE) at different doping concentrations. In the absence of dopants, the thin film of 5QLs' Sb$_2$Te$_3$ hosts a massless Dirac fermion, manifested as the hallmark linear Dirac dispersion near $\Gamma$ point~\cite{Sb2Te3} [see Fig.~\ref{Fig1}(c)].  When N-dopants are introduced, the Dirac fermion acquires a finite mass to open up a surface band gap, due to the presence of induced ferromagnetism. This satisfies the prerequisites to form QAHE~\cite{TopologicalInsulator2}.

\begin{figure}
  \includegraphics[width=8.5cm,angle=0]{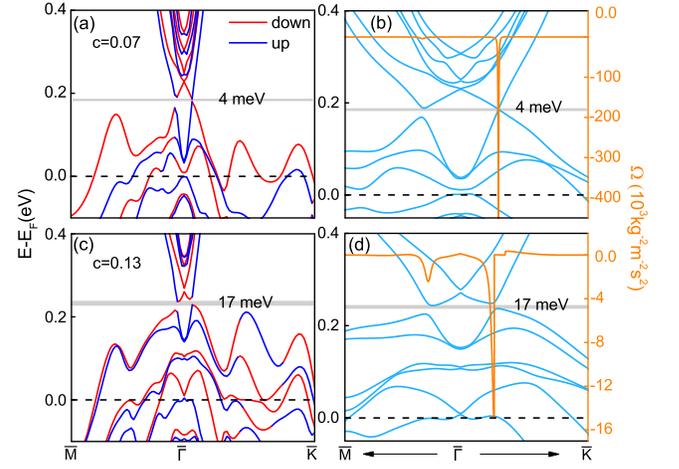}
  \caption{Band structures of N-doped Sb$_2$Te$_3$ thin film with a thickness of 5 QLs at 7\% (a,b) and 13\% (c,d) N-doping concentrations, respectively. Spin-up and -down bands are respectively highlighted in red and blue. (b) and (d): Zooming-in of the band structures near the $\Gamma$ point in panels (a) and (c) and corresponding Berry curvatures.}
  \label{Fig4}
\end{figure}

Figure~\ref{Fig4} displays the band structures and Berry curvatures along high-symmetry line near $\Gamma$ point at two different N-doping concentrations of 7\% (a,b) and 13\% (c,d), respectively. One can find that the band gap increases with the doping concentration. It is noteworthy that our calculated band structure for 7\% doping without spin opens a small gap about 7 meV due to the inter-surface hybridization. When the spin degree of freedom is invoked, the strong ferromagnetic exchange field closes the band gap and forms band crossings between spin-up and -down bands, which then naturally open up topologically non-trivial band gap when the spin-orbit coupling is considered.

Quantitatively, one can provide a unambiguous evidence for the manifestation of QAHE by integrating the Berry curvature of the occupied valence~\cite{BerryCurvature1,BerryCurvature2}. Figures~\ref{Fig4}(b) and \ref{Fig4}(d) display the Berry curvature distribution along high symmetry lines, which exhibits a large negative peak near $\Gamma$ point and zero elsewhere. As a consequence, the total integration of the Berry curvatures (i.e., the Hall conductance) must be nonzero. On the other hand, the region with finite Berry curvatures broadens with the increase of the surface band gap, and the curvature peak gradually decreases, implicating the unchanged quantized Hall conductance. Therefore, we showed that the N-doped Sb$_2$Te$_3$ thin film can realize the QAHE.

For particular doping concentrations, the Curie temperatures are respectively $T_{\rm C}=17 (29)$ Kelvin at 11\% (22\%) N-doping concentrations, by using the magnetic coupling strengths extracted from our first-principles calculations and the mean-field theory~\cite{CurieTemperature}. This indicates that the Curie temperature might be raised with increasing the doping concentration. But one has to note that higher doping concentration may decrease the spin-orbit coupling of the whole system.

One merit of N-doping in TIs is that \textit{nonmagnetic} doping scheme does not induce inhomogeneous distribution that occurs in magnetic dopants. Another one is that dissipative conduction channels from metallization in magnetically doped TIs cannot be formed by nonmagnetic and nonmetallic N doping. Since there were many materials by nonmagnetic dopants exhibiting robust high-temperature ferromagnetism ~\cite{CZnO,CZnO2,CBN,Antimonene}, our work will provide a highly desirable scheme to overcome the difficulty of observing QAHE at high temperatures~\cite{ChangCuiZu,ExperimentalQAHE1,ExperimentalQAHE2,2015NatMat}.

In conclusion, our findings demonstrate that N-doped TIs provides a versatile route to realizing high-temperature QAHE. Our proof-of-principle demonstrations revealed that nonmagnetic N-doping results in the formation of bulk gaps, long-range ferromagnetism with moderate Curie temperatures, and non-zero Berry curvatures. The estimated QAHE observation temperature is 17-29 K in 11-22\% N-doped Sb$_2$Te$_3$ thin films, which is about two orders of magnitude higher than what was previously reported. By further increasing and tailoring the doping and TI environments, it is feasible to further increase the QAHE observation temperature.

\begin{acknowledgments}
This work was financially supported by the National Key R\&D Program (2017YFB0405703), NNSFC (61434002 and 11474265), Natural Science Foundation of Hebei Province (A2019205037), Science Foundation of Hebei Normal University (2019B08), and Anhui Initiative in Quantum Information Technologies. We thank the supercomputing service of AMHPC and the Supercomputing Center of USTC for providing high-performance computing resources.
\end{acknowledgments}

\end{document}